\documentclass[epsfig,usenatbib]{mn2e}
\usepackage{graphicx}
\usepackage{multirow}

\usepackage{url}
\begin{document}

\title[Neutrino masses from  DES and Planck] {Forecasting
   neutrino masses from  galaxy clustering in the Dark Energy Survey combined with  the Planck Measurements}

\author[Lahav et al.]{Ofer Lahav$^{1}$\footnotemark,  
Angeliki Kiakotou$^1$,
Filipe B. Abdalla$^1$,
   Chris Blake$^2$ \\ \\ $^1$ Department of Physics \& Astronomy,
   University College London, Gower Street, London, WC1E 6BT, U.K.\\
   \\ $^2$ Centre for Astrophysics \& Supercomputing, Swinburne University of Technology, P.O. Box 218, Hawthorn, VIC 3122, Australia\\ \\}
   \maketitle

\begin{abstract}
We study the prospects for detecting neutrino masses from the galaxy
angular power spectrum in photometric redshift shells of the Dark Energy Survey
(DES) over a volume of $\sim 20\, h^{-3}$ Gpc$^3$, combined with the
 Cosmic Microwave Background (CMB) angular fluctuations expected to be measured from the
Planck satellite. We find that  for a $\Lambda$-CDM concordance model with 7 free parameters in
addition to a fiducial neutrino mass of  $M_{\nu} = 0.24$ eV, we recover 
from DES\&Planck  the correct value with uncertainty of $ \pm 0.12$ eV (95 \% CL), assuming perfect knowledge of the galaxy biasing. 
If the fiducial total mass is close to zero, then the upper limit is $0.11$ eV(95 \% CL).
This upper limit  from DES\&Planck is over 3 times tighter than 
using Planck alone, as DES breaks the parameter degeneracies in a CMB-only analysis.
The analysis utlilizes spherical harmonics up to 300, averaged in bin of 10 to mimic the DES sky coverage.
 The results  are similar if we supplement DES bands (grizY) with the
VISTA Hemisphere Survey (VHS) near infrared band (JHK).
The result is robust  to uncertainties in non-linear fluctuations and redshift distortions. 
However, the result is sensitive to the assumed galaxy biasing schemes and it requires 
accurate prior knowledge of the biasing.
To summarize, if the total neutrino mass in nature greater than 0.1eV, we should be able to detect it 
with DES\&Planck, a result with great importance to fundamental Physics.  

\end{abstract}
\begin{keywords}
large-scale structure of Universe -- cosmological parameters -- surveys
\end{keywords}

\section{Introduction}
\renewcommand{\thefootnote}{\fnsymbol{footnote}}
\setcounter{footnote}{1}
\footnotetext{E-mail: lahav@star.ucl.ac.uk}

Neutrinos are so far the only dark matter candidates that we actually
know exist.  It is now established from  solar, atmospheric, reactor
and accelerator neutrino experiments that neutrinos have non-zero
mass, but their absolute masses are still unknown.  Cosmology could
provide an upper limit on the sum of neutrino masses 
\citep[for review see e.g.][]{elg05,lesg06}.  
The growth of Fourier modes with comoving wavenumber 
$k > k_{\rm nr}$ will be suppressed because of neutrino free-streaming, where
\begin{equation}
k_{\rm nr} = 0.026\left(\frac{m_{\nu}}{1\;{\rm eV}}\right)^{1/2}
\Omega_{\rm m}^{1/2}\;h\,{\rm Mpc}^{-1},
\label{eq:knr}
\end{equation}
for three equal-mass neutrinos, each with mass $m_\nu$.

The current mass upper limit, obtained
using Cosmic Microwave Background (CMB) WMAP5 data, SN Ia and the BAO from 2dFGRS and SDSS, is
$M_{\nu} \equiv \sum m_{\nu} < 0.61$ eV at 95\% CL \citep{koma09}.  The challenge now is to bring down reliably the upper
limits to the 0.1 eV level or even detect the neutrino
mass.  In this way Cosmology could resolve the mass scale of neutrinos.
The new generation of deep wide surveys can play a key role in setting a
tight upper limit on the neutrino mass, and possibly detect it if the
true neutrino mass is sufficiently high.  
This is due to the order of magnitude increase in volume of the new surveys.

Here we study specifically the ability to set an upper limit on the
neutrino mass from the galaxy clustering expected in the photometric redshift survey  Dark Energy Survey (DES),
combined with Planck CMB  measurements.  The reason a survey like DES would be
effective is its large volume, $\sim 20\, h^{-3}$ Gpc$^3$, and large
number of galaxies, $\sim 300$ million.

Crudely,  for a spectroscopic survey, where accurate redshifts are known, 
the error on the power spectrum scales with the survey effective volume $V_{\rm eff}$ as
\begin{equation}
\Delta P(k)/P(k) \propto 1/{\sqrt {V_{\rm eff} }}.
\label{eq:dpkveff}
\end{equation}
On the other hand, the suppression is proportional in the linear regime
to $f_{\nu} = \Omega_{\nu}/\Omega_m$ \citep{hu98,kia08}, 
where
\begin{equation}
\Omega_{\nu} = 
\frac{\Sigma_i m_i}{93.14 h^2 ~ \rm eV}.
\label{eq:onumass}
\end{equation}
We expect therefore (when all other cosmological parameters are fixed) 
that the determination of  the upper limit on the neutrino mass
would be inversely proportional to ${\sqrt  {V_{\rm eff} } }$.
From the 2dF Galaxy spectroscopic redshift survey, covering a volume of roughly 
0.2 $(h^{-1} Gpc)^3$, the upper limit on the sum 
of neutrino mass is about 2 eV  at 95 \% CL \citep{elg02}.
Had DES been a spectroscopic  survey
with volume of about 20  $(h^{-1} Gpc)^3$, 
i.e. 100 times larger,  we would expect an upper limit of 
0.2 eV on the sum of neutrino mass.
Our detailed calculation below yields an upper limit of 0.1eV for DES\&Planck.
This is tighter than the above back-of-the-envelope calculation, probably as 
Planck priors are incorporated, and the effective volumes above are only given crudely.

However, DES is photometric redshift survey, so the radial component of distance to galaxies is significantly 
degraded, resulting in a poorer estimate of the power spectrum  \citep[e.g.][]{blk05}.
Therefore we prefer in this analysis to quantify the galaxy clustering as angular (spherical harmonic) $C_{\rm \ell}$ power spectrum 
derived in photometric redshift shells which are wide enough relative to the photometric redshift errors, and to derive the resulting upper limits more 
carefully and quantitatively.
We defer the  comparison of $P(k)$ and $C_{\rm \ell}$ approaches to future studies. 
The utility of photometric redshifts is now well-established, with many successful
techniques being employed  \citep[e.g.][]{coll04,abd09}. 
The cosmological  parameter constraints resulting from future photometric redshift
imaging surveys have been simulated by  several authors \citep[e.g.][]{seo03,dol04,zhan06}.
Application to data such as the SDSS LRG samples were given by \cite{blk07} and 
\cite{pad07}. These studies mainly emphasized 
the detection of  baryon acoustic oscillations
in the galaxy clustering pattern. Apart from the specific application to DES,  the present paper illustrates more generally  the determination of neutrino mass from photo-z surveys, to our knowledge for the first time.

This paper is organized as follows.  
In Sections \ref{sec:galsur} we summarize the DES and VHS surveys and Planck, in Section \ref{sec:photz} we present the  
photometric redshifts for DES and DES\& the Vista Hemisphere Survey (VHS) combined filters.
Sections \ref{sec:clform} give the  formalism for the galaxy angular power spectrum, and the associated joint likelihood with Planck. Section \ref{sec:result} presents the results for the basic observational and theoretical scenarios, while
Section \ref{sec:analys} provides extensions of the analysis.
An overall discussion is given in Section \ref{sec:conclu}.

\section{The Galaxy Surveys}
\label{sec:galsur}
\subsection{The Dark Energy Survey (DES)}

The Dark Energy Survey (\url{www.darkenergysurvey.org}) is a ground-based photometric survey
that will image 5000deg$^2$ of the South Galactic Cap in the optical
$griz$ bands as well as the $Y$-band. The survey will be carried out
using the Blanco 4-m telescope at the Cerro Tololo Inter-American
Observatory (CTIO) in Chile. The main objectives of the survey are to
extract information on the nature and density of dark energy and dark
matter using galaxy clusters, galaxy power spectrum measurements, weak
lensing studies and a supernova survey. This will be achieved by
measuring redshifts of some 300 million galaxies in the redshift range
$0<z<2$, tens of thousands of clusters in the redshift range $ 0 < z < 1.1$
and about 2000 Type 1a supernovae out to redshift $z \approx 1$.
Observations will be carried out over 525
nights spread over five years between 2011 and 2016. 
The DES volume is estimated  to be $23.7h^{-3}Gpc^3$ in
the range $0<z<2$  
assuming a 10$\sigma$ AB magnitude $r <   24$ \citep{manda08}.

The DES survey area overlaps with that of several other important
current and future surveys for example the southern equatorial strip
of the Sloan Digital Sky Survey and the South Pole Telescope SZE
cluster survey. 

\subsection{The Vista Hemisphere Survey (VHS)}

The entire DES region will also be imaged in the near
infra-red bands on two public surveys being conducted on the Visible
and Infra-Red Survey Telescope for Astronomy (VISTA) at ESO's Cerro
Paranal Observatory in Chile.
In particular, the Vista Hemisphere Survey (VHS)   (\url{www.vista.ac.uk})
 will image the entire southern sky ($\sim$20000deg$^2$) in the
near infra-red $YJHK_s$ bands when combined with other public
surveys. About 40\% of the total VHS time has been dedicated to
VHS-DES, a 4500deg$^2$ survey being carried out in the DES region over
125 nights in order to complement the DES optical data with near
infra-red data. The initial proposal is for the survey to image in the
$JHK_s$ bands with 120s exposure times in each band reaching
$10\sigma$ magnitude limits of $J=20.4$,$H=20.0$ and $K_s=19.4$. A
second pass may then be obtained with 240s exposures in each of the
three NIR filters in order to reach the full-depth required by
DES. The VHS-DES survey assumes that $Y$-band photometry will come
from the Dark Energy Survey.

\subsection{Planck}

The recently launched ESA's 
Planck satellite will map the CMB with better resolution, sensitivity 
and frequency (in nine bands from one centimeter to one third of a millimeter) 
than previous CMB experiments \footnote{\url{http://www.esa.int/SPECIALS/Planck}}.

We have obtained forecasts from the Planck satellite by using the
technique described in \citep{abd07}. 
We have assumed conservative values for the sensitivity by
taking only one usable science channel with 8 detectors, a noise
effective temperature of $120 \mu K \sqrt{s}$, an angular resolution
of 10 arcminutes, and 65\% sky coverage in a one year survey.  The input fiducial
masses are taken as $\Omega_{\nu}=0.001$  and 0.005,
and the other cosmological
parameters considered as listed in Table \ref{table:fiducial}.

\begin{table*}
 \begin{tabular}{|c|c|c|c|c|c|c|c|}
 \hline
 $\Omega_{\nu}$ & $\Omega_c$ & $\Omega_b$ & $h$ & $\sigma_8$  & $n_s$ & $\tau$ & $N_{\nu}$\\
 \hline
 0.005 &  0.255 &  0.04 & 0.72 & 0.9 & 1 & 0.166 & 2.0\\
 0.001 &  0.259 &  0.04 & 0.72 & 0.9 & 1 & 0.166 & 2.0\\
 \hline
  \end{tabular}
\caption{The two fiducial models in the Planck simulations. All these 8 parameters vary in the Planck MCMC chain.
 A  flat universe is assumed.  $N_{\nu}$ is the number of  massive neutrinos.  We note the value of the optical depth $\tau$ assumed here is based on WMAP1, higher than the most recent value estimated from WMAP5. }
\label{table:fiducial}
\end{table*}

\section{Photometric redshifts}
\label{sec:photz}

We determined photometric redshift estimates as described in \cite{manda08} using mock samples of DES and VHS and the photo-z software
package ``{\tt ANNz}'' \citep{coll04}.  In brief,
artificial neural networks are applied to parameterize  a non-linear relation
between the galaxy redshift and the
galaxy multi-band photometry. The neural network is ``trained'' using a set of galaxies
with known true redshifts by minimizing  the sum of the squared differences between the
photometric and true redshifts.  For comparison of ANNz with 5 other photoz codes see
\cite{abd09}.

\begin{figure}
\includegraphics[width=8cm,height=5cm]{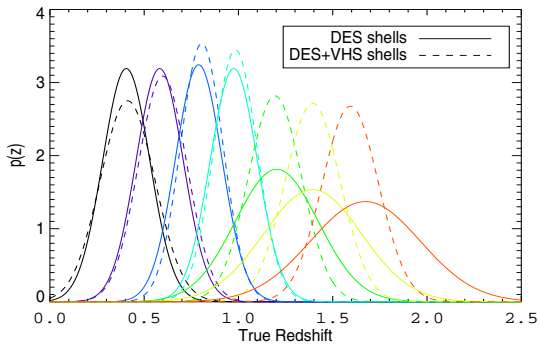}
\caption{Gaussian fits to the true redshift distribution of
  in each of seven photometric redshift slices
  analyzed in this study.  
The best-fitting Gaussian parameters $\mu$
  and $\sigma$ are indicated, where $p(z) \propto
  \exp{\{-[(z-\mu)^2/2\sigma^2]\}}$.}  
\label{fig:zshells}  
\end{figure}

Figure \ref{fig:zshells} shows the distribution of true
redshifts in the evaluation set for galaxies binned in seven
photometric-redshift slices, as given in Table \ref{table:zshells}.
We find that these error distributions are well fitted
by Gaussian distributions; the best-fitting values of the mean and
standard deviation for the different slices are indicated in the
Table.  These Gaussian functions are taken as our model for the
redshift distribution of galaxies in each photo-$z$ slice when
analyzing the galaxy clustering results.  

\begin{table*}
 \begin{tabular}{|c|c|c|c|c|c|}
 \hline
 photo-z shell& galaxy &\multicolumn{2}{ | c | }{ DES } & \multicolumn{2}{|c|}{DES\&VHS} \\
 &  fraction & mean & $\sigma$ & mean & $\sigma$\\
 \hline
 0.3$<z_{ph}<$0.5 & 0.211 & 0.405 & 0.125 & 0.412 & 0.145\\
 0.5$<z_{ph}<$0.7 & 0.337 & 0.582 & 0.125 & 0.598 & 0.129\\
 0.7$<z_{ph}<$0.9 & 0.215 & 0.789 & 0.123 & 0.805 & 0.113\\
 0.9$<z_{ph}<$1.1 & 0.128 & 0.975 & 0.125 & 0.984 & 0.116\\
 1.1$<z_{ph}<$1.3 & 0.098 & 1.203 & 0.220 & 1.193 & 0.142\\
 1.3$<z_{ph}<$1.5 & 0.081 & 1.393 & 0.260 & 1.393 & 0.147\\
 1.5$<z_{ph}<$1.7 & 0.027 & 1.673 & 0.291 & 1.593 & 0.149\\
\hline
\end{tabular}
\caption{ The best-fitting Gaussian parameters $\mu$
  and $\sigma$, in true redshift $z$,  where $p(z) \propto
  \exp{\{-[(z-\mu)^2/2\sigma^2]\}}$ for photo-z shells derived
from DES alone (5 filters) and DES\&VHS (8 filters).}.  
\label{table:zshells}
\end{table*}

\section{The angular power spectrum in photo-z shells}
\label{sec:clform}

\subsection{Spherical Harmonic Formalism}
\label{secclest}

It is common to expand the distribution of galaxies in spherical harmonic 
coefficients $a_{\ell,m}$,
which are then averaged to form the angular power spectrum $C_\ell$
\begin{equation}
<|a_{\ell,m}|^2> = C_\ell \, .
\label{eq:alm}
\end{equation}
\citep[see e.g.][]{peeb73,sch92,wri94,wand01}.  

The angular power spectrum $C_\ell$ is a projection of the spatial
power spectrum of fluctuations at different redshifts $z$, $P(k,z)$,
where $k$ is a co-moving wavenumber.  
When the sky coverage is incomplete 
the prediction for observed angular power spectrum
can be estimated through convolution, as explained later in the paper.

We follow here the notation of  \citep{blk07}. 
The equation for the projection
is:
\begin{equation}
C_\ell = \frac{2 \, b^2}{\pi} \int P_0(k) \, g_\ell(k)^2 \, dk
\label{eq:clexact}
\end{equation}
where we assume $P(k,z) = P_0(k) \, D(z)^2$, with $D(z)$ the linear growth
factor at redshift $z$, and $g_\ell(k)$ contains $D(z)$ as defined below.

We note that this decomposition of $P(k,z)$
is strictly only valid in linear theory, and its application at
smaller scales is an approximation. Moreover, 
in the presence of massive neutrinos $P(k,z)$ cannot be decomposed 
even in linear theory.
However, we find that over the redshift  $z<2$  and $k$-range of interest 
 $P(k,z) \approx P_0(k)  D(z)^2$
to within 0.5\%
\citep[cf.][]{lesg06,kia08}.
We have also assumed in the above equation a
scale-independent and epoch-independent bias factor $b$.
Later we allow the biasing to vary with redshift.

\begin{figure}
\center
\includegraphics[width=8cm,height=6cm]{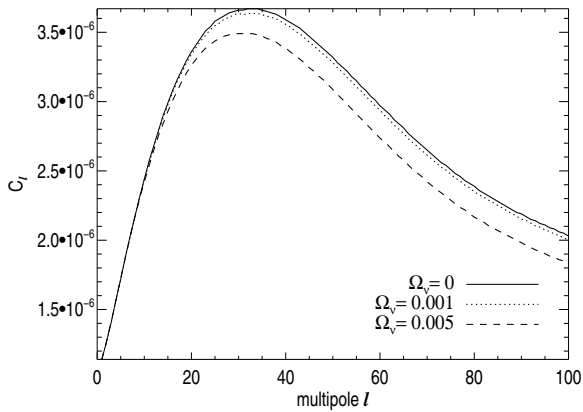}
\caption{The model angular power spectrum for the photometric redshift
  slice $z =0.8$ with  the exact
  expression in real space (equation \ref{eq:clexact} with the kernel
  of equation \ref{eq:ker}). The curves are for three values of $\Omega_{\nu} =0.000, 0.001$ and
  $0.005$, with the other cosmological parameters held fixed 
  at the values given  in Table \ref{table:fiducial}.
  }
\label{fig:sh3cl}
\end{figure}

\begin{figure}
\center
\includegraphics[width=8cm,height=6cm]{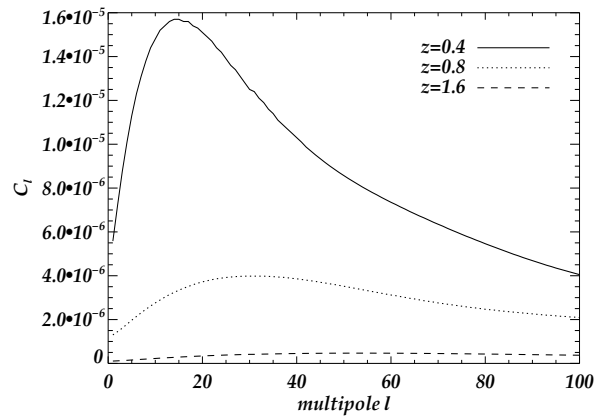}
\caption{ The model angular power spectrum for three photometric redshift
 slices centered at $z=0.4, 0.8$ and $1.6$.
 The curves are for  $\Omega_{\nu} =0.005$,
 with the other cosmological parameters held fixed 
 at the values given in the first row Table \ref{table:fiducial}. The amplitude difference between these curves is partially due to the growth of structure with redshift and partially due to the variation in the photo-z error with   redshift.}
 \label{fig:nu05cl}
\end{figure}

Figure \ref{fig:sh3cl} shows the expected spherical harmonic $C_{\ell}$ (ignoring shot noise) for three values of assumed neutrino mass,
illustrating that the suppression effect can  be measured. 
Figure \ref{fig:nu05cl} shows the expected  $C_{\ell}$'s for three photo-z shells.  

Similarly, $C^{i,j}_\ell$ is the cross angular power spectrum between the
two redshift slices $i$ and $j$ , with 
a kernel $g^i_\ell(k)$ for a redshift slice $i$:
\begin{equation}
C^{i,j}_\ell = \frac{2 \, b_i \, b_j}{\pi} \int P_0(k) \, g^i_\ell(k)
\, g^j_\ell(k) \, dk
\label{eq:clcross}
\end{equation}
where $b_i$ and $b_j$ are the linear bias factors for the slices.
The kernel $g_\ell(k)$ is given by:
\begin{equation}
g_\ell(k) = \int_0^\infty j_\ell(u) \, f(u/k) \, du \, .
\label{eq:ker}
\end{equation}
Here, $j_\ell(x)$ is the spherical Bessel function and $f(x)$ depends
on the radial distribution of the sources as
\begin{equation}
f[z(x)] = p(z) \, D(z) \left( \frac{dx}{dz} \right)^{-1}
\end{equation}
where $x(z)$ is the co-moving radial co-ordinate at redshift $z$, and
$p(z)$ is the redshift probability distribution of the sources,
normalized such that $\int p(z) \, dz = 1$.
A good
approximation for  equation \ref{eq:clexact} which is valid for moderately large
$\ell \ga 30$) is:
\begin{equation}
C_\ell = b^2 \int P_0(k=\ell/x) \, D^2(z) \,  x(z)^{-2} \, p(z)^2 \, \left(
\frac{dx}{dz} \right)^{-1} dz \, .
\label{eq:clapprox}
\end{equation}
The above formalism neglects redshift distortions,
which we examine later.

The cosmological
parameters enter the matter power spectrum of fluctuations $P_0(k)$, the growth factor $D(z)$
and the co-moving distance $x(z)$ in the above equations.  For the
redshift distribution of the sources $p(z)$ we used the Gaussian
functions fitted to the training set data (see Figure \ref{fig:zshells}).
We derived model spatial power spectra using the ``{\tt CAMB}''
software package \citep{lew00}.
For more technical details of these calculations see \citep{blk07}.

\subsection{Spherical Harmonic Likelihood}

In this work we forecast the measurements, without any data, so we use
the following relation to express the ratio of (log) probabilities
between a selected model in the parameter space B and the assumed fiducial model A. 
The model describes all the power spectra produced from 
random Gaussian fields with a certain degree of cross correlation 
\citep{buch02,abd07}.

\begin{eqnarray}
\begin{array}{ll}
\displaystyle{{\rm log} \left( \frac{p_B}{p_A}\right)} & = 
\displaystyle{{\rm log} \left(\frac{p(a_{lm}|B)}{p(a_{lm}|A)}\right)}\\
\,  & = \displaystyle{\frac{f_{sky}}{2}\sum_l (2l+1) 
\left( Tr(I - M_A M^{-1}_B) + \right.} \\
\,  &  \,\,\,\,\,\,\,\,\,\,\,\,\,\,\,\,\,\,\,\,\,\,\,\,\,\,\,\,\,\,\,\,\,\,\,\,\,\,\,\,\,\,\,\,\,\,\,\,\,\,\,\,\,\displaystyle{\left.{\rm log}\left(Det(M_A M^{-1}_B)\right)\right)}
\end{array}
\label{eq:likecross}
\end{eqnarray}
where $I$ is the identity matrix and
the matrices $M_A$ and $M_B$ are given by the values of 
the individual power spectra and their cross correlations at a given 
mode $l$. As in our case we have 7 shells 
each of them is a $7 \times 7$ matrix.   
The diagonal terms are
\begin{equation}
{\rm diag}  (M_l) = C_l + \sigma_l^2,
\label{eq:diag}
\end{equation}
\noindent
where $\sigma^2_l = \frac{1}{N/\Delta\Omega}$
where $N/\Delta\Omega$ is the average source density 
and $f_{\rm sky} =  \Delta \Omega /(4 \pi) $ denotes the fraction of sky covered by
the survey.   There is no shot-noise in the off-diagonal terms. 

To get insight into  this likelihood ratio we note the special  case of eq. 
\ref{eq:likecross}
for a shell's  auto-correlation \citep[cf.][]{fish94}: 
\begin{eqnarray}
\begin{array}{ll}
\displaystyle{{\rm log} \left( \frac{p_B}{p_A}\right)} & = 
\displaystyle{{\rm log} \left(\frac{p(a_{lm}|B)}{p(a_{lm}|A)}\right)}\\
\,  & = \displaystyle{\frac{f_{sky}}{2}\sum_l (2l+1) 
\left( 1- \frac{C_{l,A}+\sigma_l^2}{C_{l,B}+\sigma_l^2}\right.} + \\
\,  & \,\,\,\,\,\,\,\,\,\,\,\,\,\,\,\,\,\,\,\,\,\,\,\,\,\,\,\,\,\,\,\,\,\,\,\,\,\,\,\,\,\,\,\,\,\,\,\,\,\,\,\,\,\,\,\,\,\,\,\displaystyle{\left.{\rm log}\left(\frac{C_{l,A}+\sigma_l^2}{C_{l,B}+ \sigma_l^2}\right)\right)}.
\end{array}
\label{eq:likeauto}
\end{eqnarray}

\section{Analysis and Results}
 \label{sec:result}
 The joint likelihood is implemented as follows. The Planck MCMC chain in 8-dimensional space
 is centred at one of the two fiducial models  given in Table \ref{table:fiducial}.
 The DES likelihood is evaluated using eq. \ref{eq:likecross}, with label  $A$ 
corresponding to the fiducial model, and label $B$  to another point of the Planck chain.
By evaluating the DES likelihood at the Planck chain points we 
effectively get the joint likelihood of DES \& Planck.  
If we are interested in a subset of the parameters of interest we can easily marginalize over the 
other parameters by projecting the distribution of the points in the resulting chain.

We first consider a baseline model so we can see ``the wood for the trees", where we  assume a linear matter power spectrum
(derived from CAMB), a linear biasing parameter $b=1$ for all 7 photo-z  slices, no redshift distortion, and we ignore the mask.
 In the discussion below we refer to  the 95\% CL values. 
 
\begin{figure}
\includegraphics[width=9cm,height=5cm]{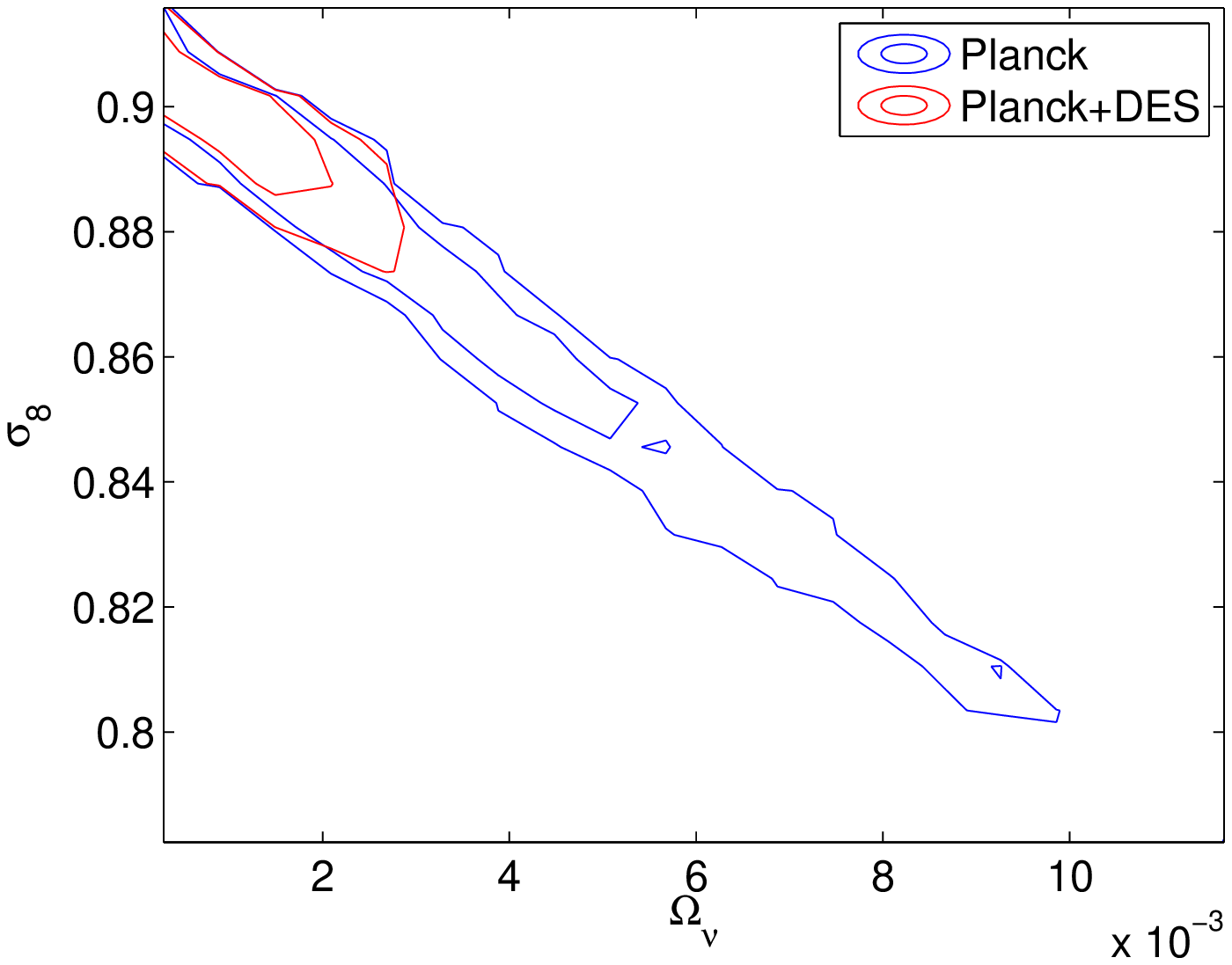}
\includegraphics[width=9cm,height=5cm]{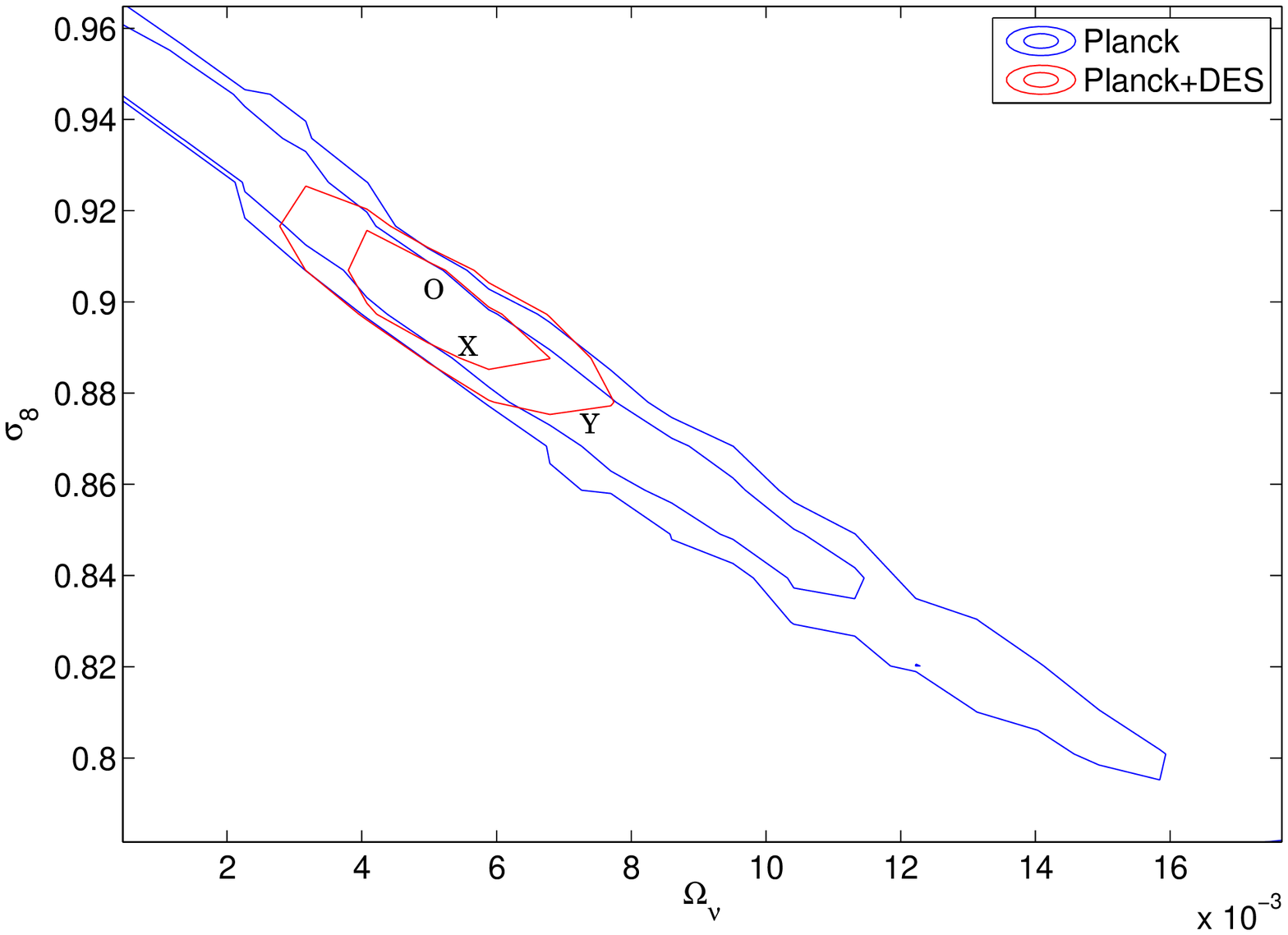}
\caption{The likelihood contours  show $\Omega_{\nu}$ and $\sigma_8$ from Planck alone and Planck \& DES ,
with fiducial  $\Omega_{\nu} =0.001$ (top) and $\Omega_{\nu} = 0.005$ (bottom). 
The likelihood functions are evaluated for $\ell_{max}=300$, where $\Delta \ell =10$ (corresponding to entry 5 in Table
\ref{table:results}.
 These plots are for seven cross-correlated DES shells. Contours show 68\% CL and 95\% CL.
 The sensitivity to the assumed galaxy biasing scheme is illustrated with  
 the point $O$ corresponds to $b=1.00$,  $X$ corresponds to $b=1.02$ 
 constant for all shells, and $Y$ corresponds to epoch-dependent $b(z)$ from eq \ref{eq:bz}
 with $b_0=1.02$.
}
\label{fig:contsig8}
\end{figure}

\begin{figure}

\includegraphics[width=9cm,height=5cm]{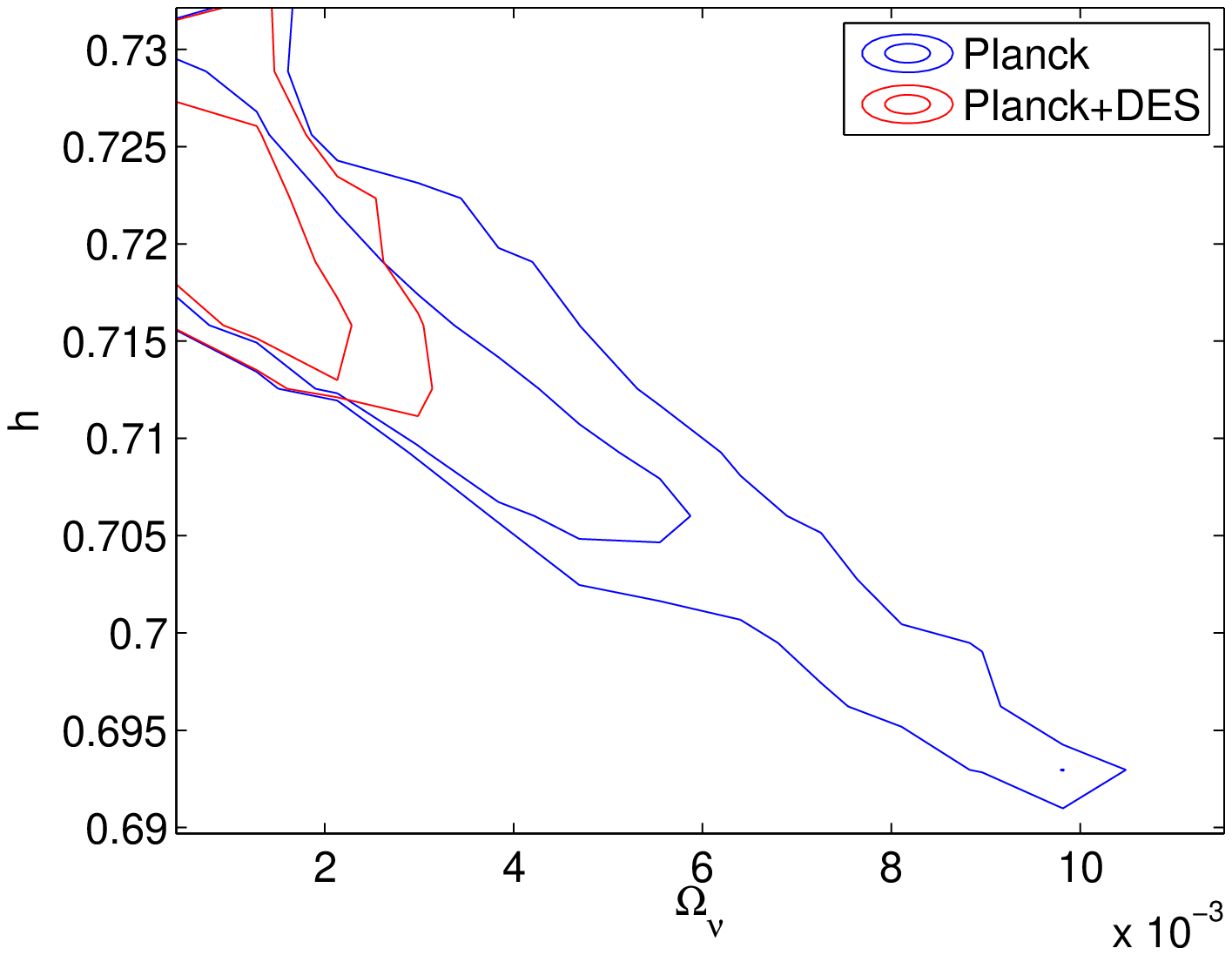}
\includegraphics[width=9cm,height=5cm]{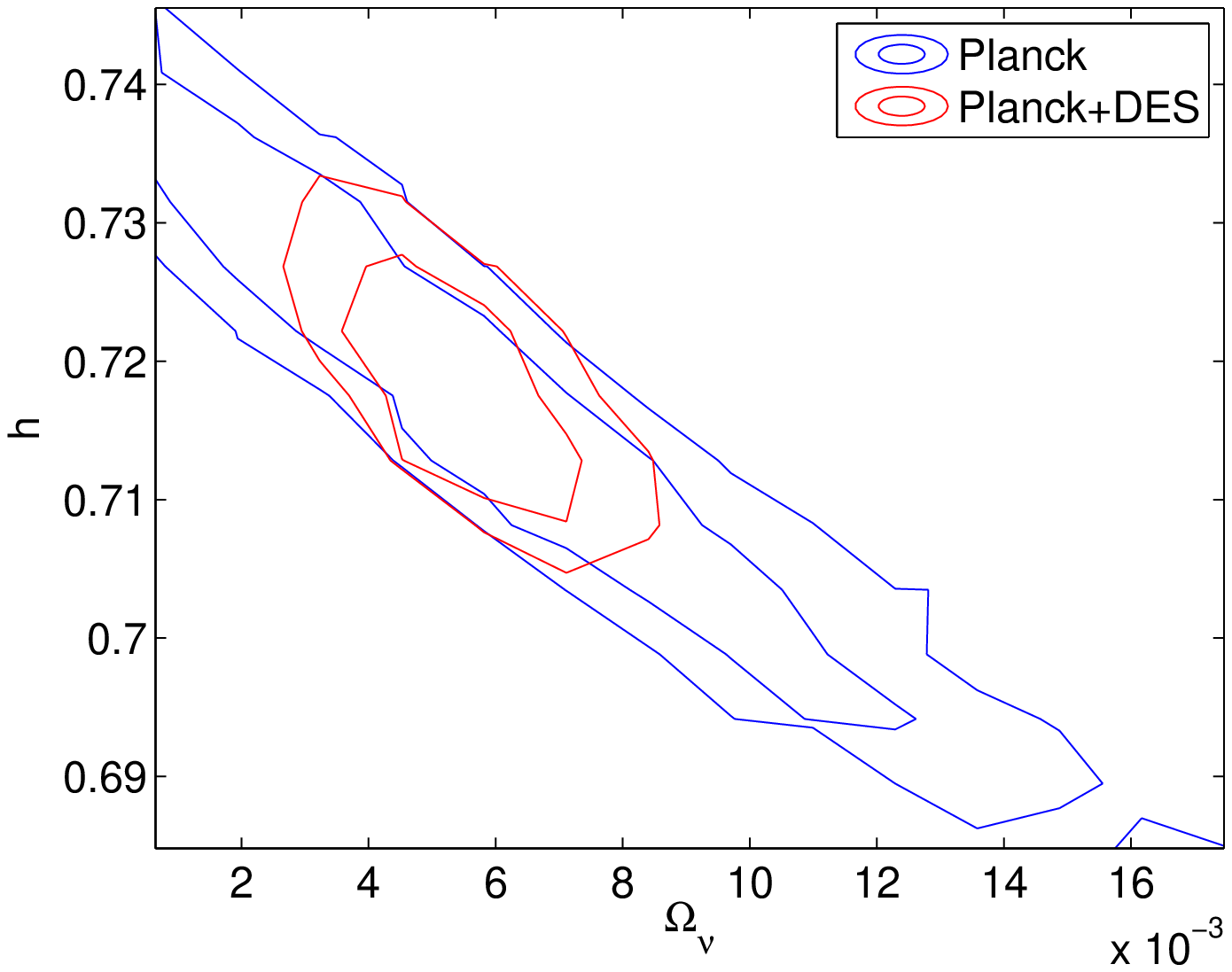}
\caption{The likelihood contours  show $\Omega_{\nu}$ and $h$ from Planck alone and Planck \& DES ,
with fiducial  $\Omega_{\nu} =0.001$ (top) and $\Omega_{\nu} = 0.005$ (bottom). 
The likelihood functions are evaluated for $\ell_{max}=300$ and $\Delta \ell =10$.
 These plots are for seven cross-correlated DES shells. Contours show 68\% CL and 95\% CL. }
\label{fig:conthh}
\end{figure}

\begin{figure}
\includegraphics[width=9cm,height=5cm]{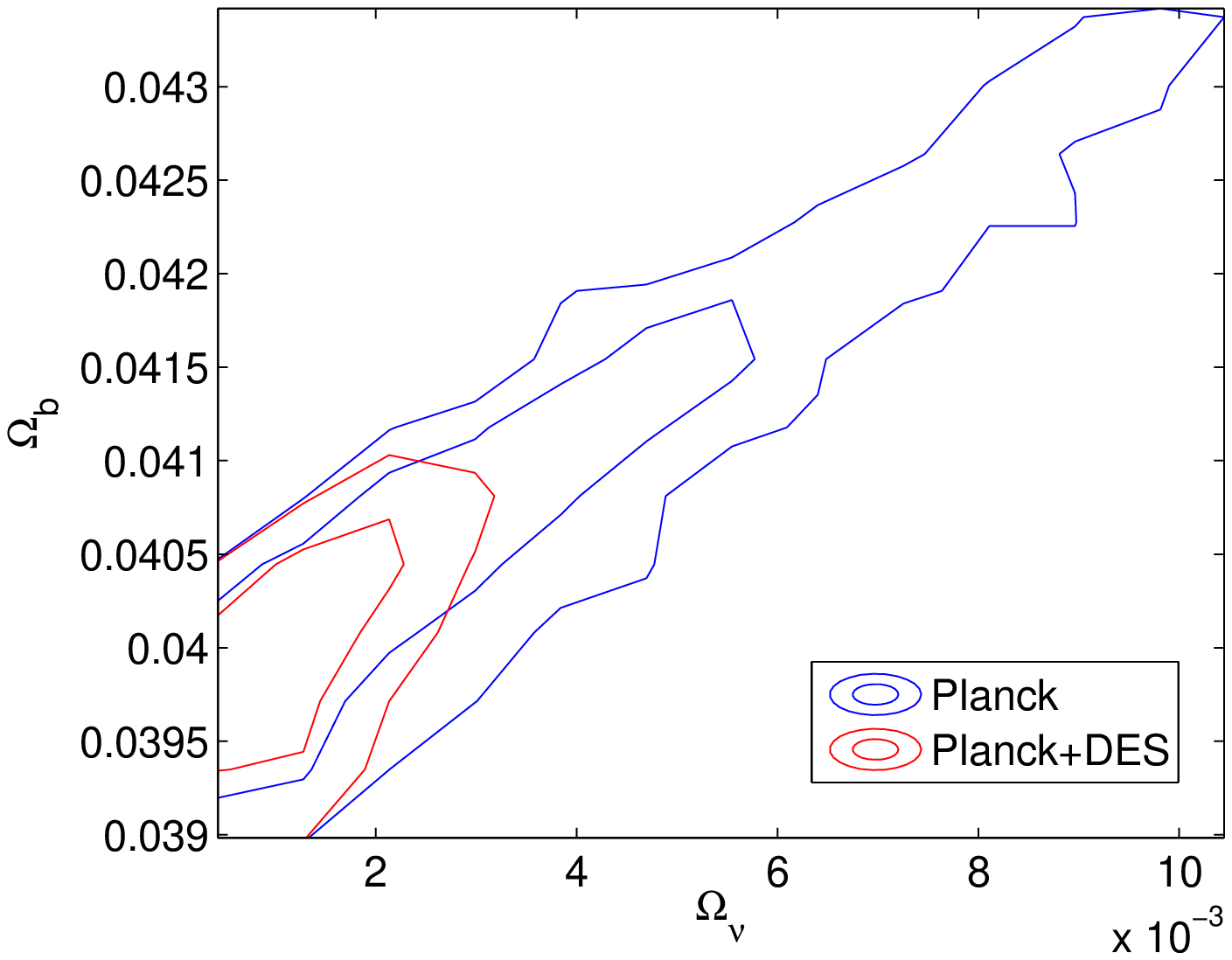}
\includegraphics[width=9cm,height=5cm]{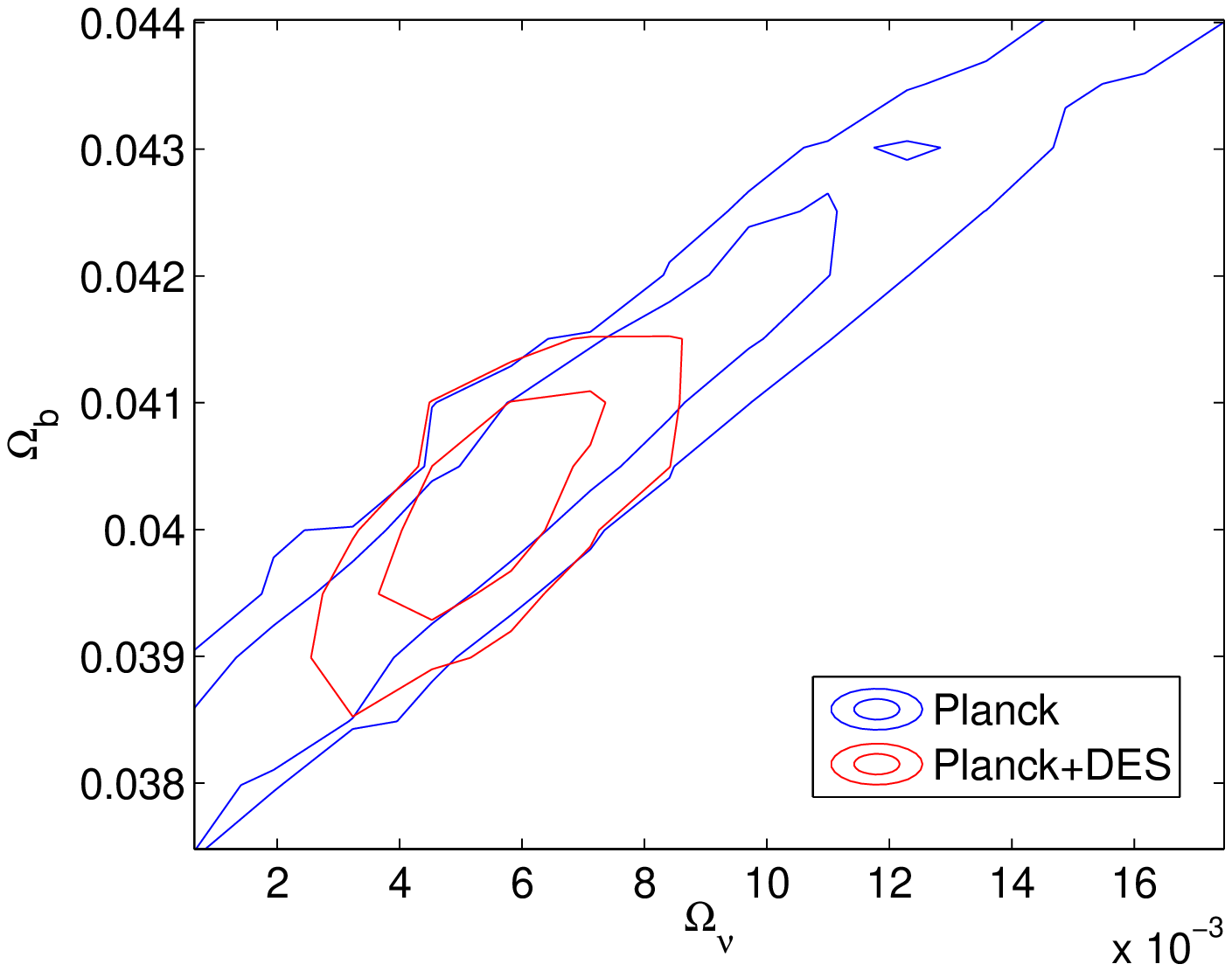}
\caption{The likelihood contours  show $\Omega_{\nu}$ and $\Omega_b$ from Planck alone and Planck \& DES ,
with fiducial  $\Omega_{\nu} =0.001$ (top) and $\Omega_{\nu} = 0.005$ (bottom). 
The likelihood functions are evaluated for $\ell_{max}=300$ and $\delta \ell = 10$.
 These plots are for seven cross-correlated DES shells. Contours show 68\% CL and 95\% CL.}
\label{fig:contob}
\end{figure}

\begin{figure}
\includegraphics[width=9cm,height=5cm]{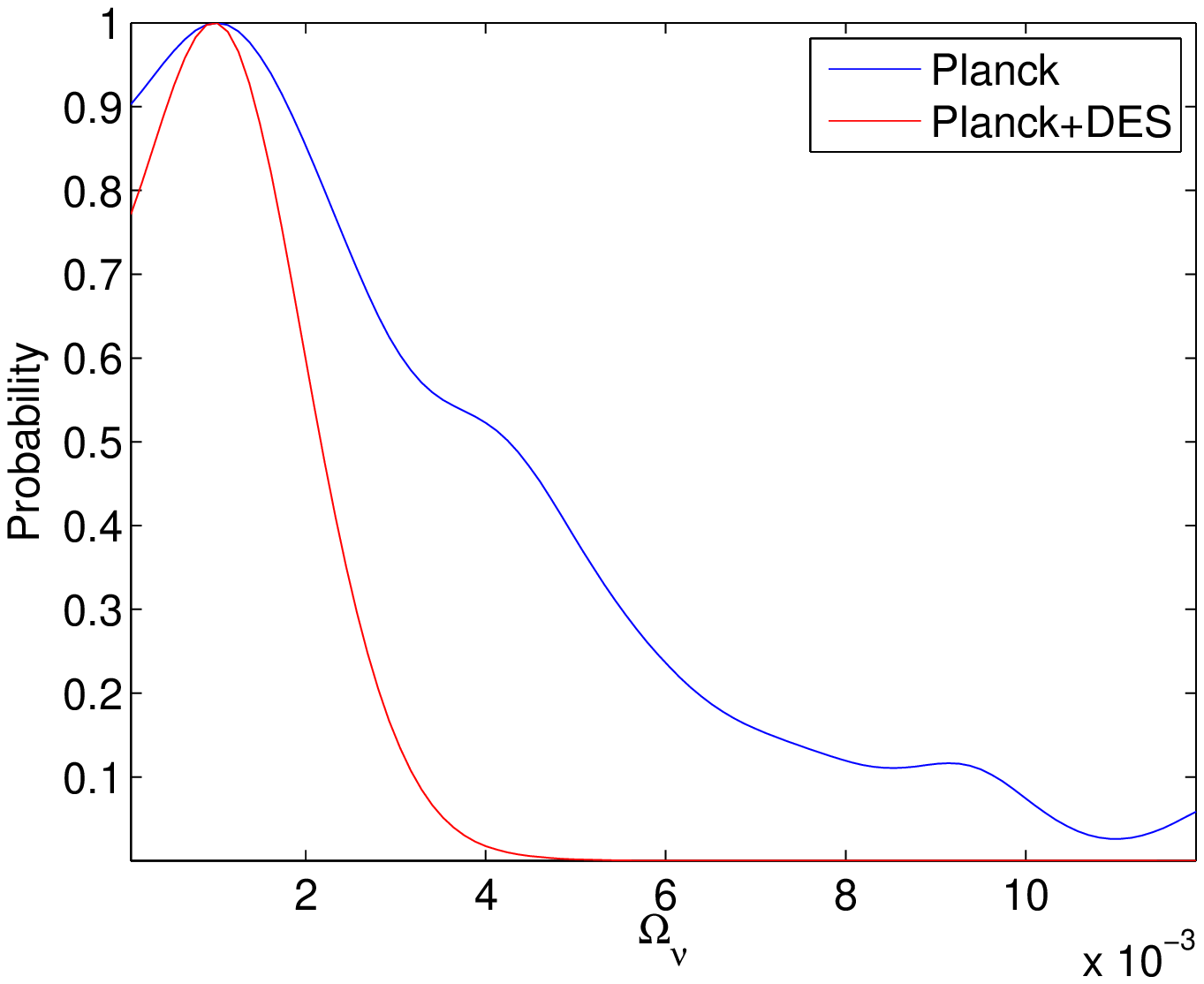}
\includegraphics[width=9cm,height=5cm]{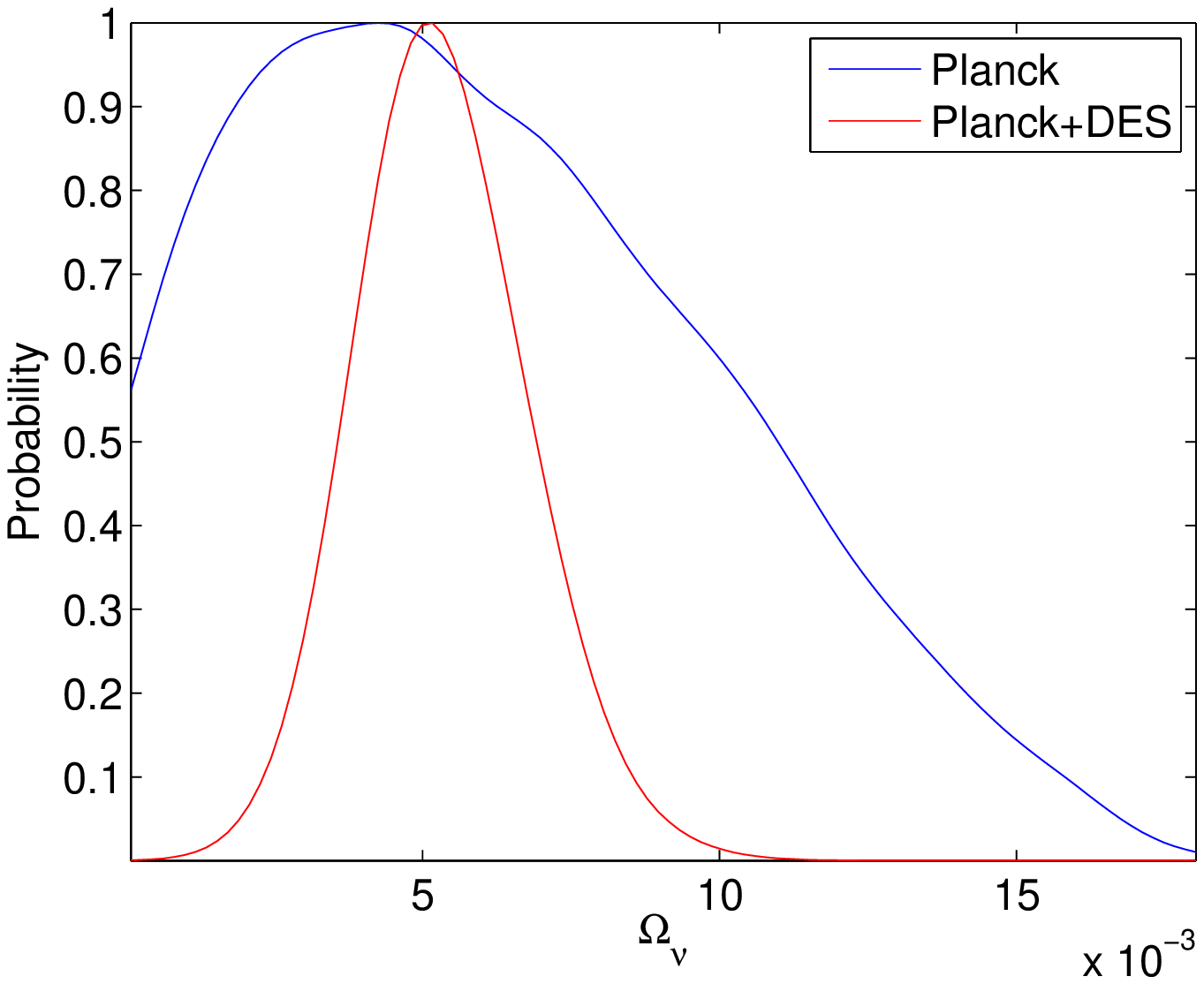}
\caption{The probability for the neutrino mass density $\Omega_{\nu} $, 
with fiducial  $\Omega_{\nu} =0.001$ and $\Omega_{\nu} = 0.005$,
after marginalizing over the other 7 free parameters,
 for the cases of Planck alone and Planck \& DES.}
\label{fig:probonu}
\end{figure}

 The first entry in Table \ref{table:results} is for Planck alone, 
 while the second entry is for Planck \& DES, both for $l=1,2...,100$.
 We see that the addition of DES improves the upper limit to 0.11 eV, more than 3 times tighter than Planck alone.
 The corresponding error bar  on the high-mass values is 0.13 eV from Planck \& DES.
The likelihood contours in Figures \ref{fig:contsig8},\ref{fig:conthh},\ref{fig:contob} correspond to entry 5 in Table \ref{table:results}, 
where $\ell_{max}=300$, where $\Delta \ell =10$,
which roughly  represents the DES incomplete sky,
with fiducial  $\Omega_{\nu} =0.001$ and $\Omega_{\nu} = 0.005$. 
They show  pairs of variables $(\Omega_{\nu}, \sigma_8)$, $(\Omega_{\nu}, h)$, and $(\Omega_{\nu}, \Omega_{\rm b})$, 
and in Figure \ref{fig:probonu} the probability for $\Omega_{\nu}$, after marginalizing over all other seven parameters. Numerical results are given in Table \ref{table:results}, as 68 \% and 95 \% upper limits. 
We note the similarity of results of entries 2 and 5 in that Table.
We also note that in the case of entry 5, the 5-$\sigma$ upper limit is 0.121 eV, 
compared with the 2-$\sigma$ value of 0.106 eV.

\begin{table*}
\begin{tabular}{|c|c|c|c|c|c|c|c|}
\hline

\multirow{2}{*} &\multirow{2}{*}{$l$min} &\multirow{2}{*}{$l$max}&\multirow{2}{*}{$\Delta l$}&\multicolumn{2}{|c|}{$M_{\nu}$=0.048 eV}&\multicolumn{2}{|c|}{$M_{\nu}$=0.241 eV} \\
\cline{5-8}
& &&&$68\% \textrm{CL}$ & $95\% \textrm{CL}$ & $68\% \textrm{CL}$ & $95\% \textrm{CL}$ \\
\hline
(1) Planck &1&100&1&$M_{\nu} <$0.175 &$M_{\nu} <$ 0.386&0.034$< M_{\nu} <$0.391&0.0041$< M_{\nu} <$0.616 \\
(2) Planck\&DES &1&100&1& $M_{\nu} <$ 0.067&$M_{\nu} <$ 0.112&0.172$< M_{\nu} <$0.304&0.113$< M_{\nu} <$0.380\\
(3) Planck\&DES&20&120&1&$M_{\nu} <$ 0.063&$M_{\nu} <$ 0.101&0.176$< M_{\nu} <$0.296&0.124$< M_{\nu} <$0.362\\
(4) Planck\&DES&1&100&10&$M_{\nu} <$ 0.106&$M_{\nu} <$ 0.216&0.109$< M_{\nu} <$0.354& 0.014$< M_{\nu} <$0.476\\
(5) Planck\&DES&1&300&10&$M_{\nu} <$ 0.063&$M_{\nu} <$0.106 &0.169$< M_{\nu} <$0.291&0.113$< M_{\nu} <$0.354\\
(6) Planck\&DES&1&300&1&$M_{\nu} <$0.052&$M_{\nu} <$0.079&0.181$< M_{\nu} <$0.275&0.136$<M_{\nu} <$0.322\\
(7) Planck\&DES\&VHS&1&100&1&$M_{\nu} <$ 0.066&$M_{\nu} <$ 0.110&0.174$< M_{\nu} <$0.304&0.117$< M_{\nu} <$0.379\\
(8) Planck\&DES\&VHS&1&300&10&$M_{\nu} <$0.061 &$M_{\nu} <$0.101 &0.171$< M_{\nu} <$0.291&0.117$< M_{\nu} <$0.353\\

\hline

\end{tabular}
\caption{Table showing derived results neutrino mass $M_{\nu}$  for two input fiducial values $M_{\nu} = 0.048$ eV and $0.241$ eV.
The results are given for different combination of mock data sets (DES, VHS, Planck) and different 
range of spherical harmonics $l_{min}, l_{max}$ and smoothing scale $\Delta l$.}.
\label{table:results}
\end{table*}

\section{Extensions of the analysis}
\label{sec:analys}
\subsection{Incomplete sky coverage}

When the sky coverage is incomplete 
the prediction for observed angular power spectrum
can be estimated through convolution: 
\begin{equation}
< C_\ell^{\rm obs} > = \sum_{\ell} R_{\ell,\ell'} \, C_{\ell'}
\label{eq:clobs}
\end{equation}
where the  ``mixing matrix'' $R_{\ell,\ell'}$ can be determined from the
angular power spectrum  of the survey window function.
For a survey of 5000 sq deg or so the mixing matrix smears out harmonics over
$\Delta \ell  \approx 10 $ , subject to survey's geometry  \citep{blk07}.
We illustrate the effect of incomplete sky coverage by averaging the angular power spectra in multipole bands of
width $\Delta \ell = 10$:
\begin{equation}
< C_\ell^{\rm av} > = \sum_{\ell'} (2 \ell' + 1)   C_{\ell'}/ \sum_{\ell'} ( 2\ell' +1)
\label{eq:clav}
\end{equation}
Entry (4) In Table \ref{table:results} shows that the upper limit on the neutrino mass is twice as large as for a `whole sky DES'
(entry 2 in Table \ref{table:results}).
However, if we use harmonics up to $\ell_{\rm max} = 300$  (entry 5), then the upper limits are very similar to those in entry (2),
where $\ell_{\rm max} = 100$.  

\subsection{DES \& VHS photoz} 

Entries (7) and (8) in Table \ref{table:results} illustrate the impact 
of adding  VHS near infrared photometry, i.e. having photo-z based on 8 filters.
It turns out that the results on neutrino mass look very similar to DES alone.
This is because the improvement of VHS is at high redshift \citep{manda08} and the high redshift shells have higher shot noise and hence are weighted less in the likelihood analysis. 

\subsection{Redshift distortion}

These distortions significantly
affect the amplitude of the projected power spectrum on large scales
$\ell \la 50$, owing to the relative narrowness of each redshift
slice.  The amplitude of the redshift-space distortions is controlled
by a parameter $\beta(z) \approx \Omega_{\rm m}(z)^{0.6}/b(z)$ \footnote{For a more accurate expression in the presence of neutrinos see \cite{kia08}, but this correction is negligible for the qualitative argument presented here.}, where
the quantities on the right-hand side of the equation are evaluated at
the centre of each redshift slice of our analysis.  The effect is to
introduce an additional term to the kernel of equation \ref{eq:ker}
such that it becomes $g_\ell(k) + g^\beta_\ell(k)$ where:
\begin{equation}
g^\beta_\ell(k) = \frac{\beta}{k} \int_0^\infty j_\ell^\prime(u) \,
f^\prime(u/k) \, du
\label{eq:kerbeta}
\end{equation}
\citep{fish94,pad07,blk07}.

\begin{figure}
\includegraphics[width=8cm,height=5cm]{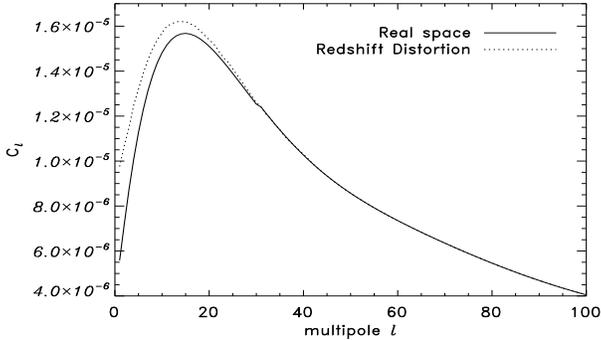}
\caption{ Real-space (bottom) vs redshift space (top) for the nearest ($z=0.4$) shell
assuming $\Omega_{\nu} =0.005$ and the other cosmological parameters as in Table \ref{table:fiducial}. The curves are normalized to agree at high $l$.
The assumed biasing is $b=1$.}  
\label{fig:clz}
\end{figure}
Figure  \ref{fig:clz} shows that the redshift distortion is important for $\ell < 20$.  
In entry 3  of Table \ref{table:results} we show the effect of analysing $20 < \ell <  120$. 
Excluding the low $\ell$ changes only little the results on neutrino mass compared with the case where low-$\ell$ are included (entry 2).  

\subsection{ Non-linear power spectrum}

Typically, non-linearity in the power spectrum appears for 
 $k > k_{\rm max} \approx  0.15 \, h$ Mpc$^{-1}$, at $z=0$  \citep[e.g.][]{smit03} and higher $k_{\rm max}$ at higher redshifts.  We can estimate the
equivalent maximum multipole $\ell_{\rm max}$ of the angular power spectrum using the 
scaling relation  $k_{\rm max}  = \ell_{\rm max} /r(z)$ (see equation \ref{eq:clapprox}).
For example,   for the  shells with mean redshift  $z \approx 0.4, 0.8$ and $1.6$  and assuming a flat universe with $\Omega_{\rm m}=0.25$ 
this corresponds to $\ell_{\rm max} \approx 160, 290$ and $510$.
Roughly speaking it is safe to assume linear theory up to those $\ell_{\rm max}$ values.

To check the justification of this simple approximation, we considered a  non-liner power spectrum 
in the presence of massive neutrinos.
The total matter density fluctuation can be written as
\begin{equation}
\delta_{\rm m} = f_{cb} \delta_{cb} + f_{\nu} \delta_{\nu}  \;, 
\end{equation}
where $f_{cb} = 1 - f_{\nu}$ is the fractional contribution  of the CDM plus baryon  of the present epoch 
mass density.
The resulting non-linear power spectrum can be approximated \citep{sait08,sait09,ichi09} as
\begin{equation}
P^{\rm nl}_m(k) = f_{\nu}^2 P_{\nu}^{L}(k) + 
(1-f_{\nu})^2 P_{cb}^{NL}(k) + 2 f_{\nu} (1-f_{\nu})P_{cb,\nu}^L(k) \;, 
\label{eq:pknl}
\end{equation}
where superscripts $L$ and $NL$ denote linear and non-linear power spectra respectively,
and $P_{cb,\nu}^L (k)$ is a cross power spectrum neutrino and CDM plus baryons.

This approximation is convenient, as the non-linear  $P_{cb}^{NL}(k)$ 
can be taken from fits to N-body simulations (\citep{sait08} based on \citep{smit03}; ``{\tt halofit = 1}'' in {\tt CAMB}).
Assuming linearity for the  power spectra which involve neutrino mass is reasonable,  
as due to their free streaming, massive neutrinos remain in the linear regime, rather than joining the non-linear evolution of CDM plus baryons.  
We also note that the pre-factor $f_{\nu}$ is small,  which justifies further ignoring non-linear neutrino perturbations. 
For a slightly different approximation see \citep{hann06}.

\begin{figure}
\includegraphics[width=8cm,height=5cm]{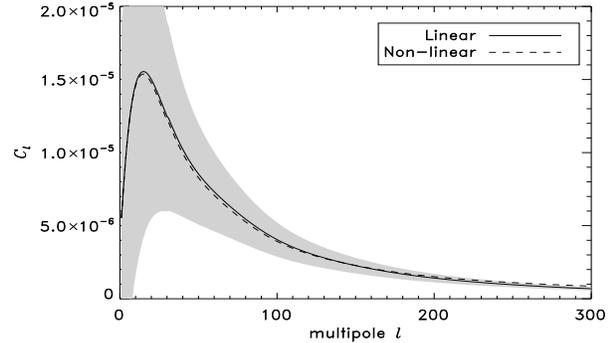}
\caption{ linear power spectrum (solid line) vs non-linear power spectrum  according to Eq. \ref{eq:pknl} (dashed line) for the nearest ($z=0.4$) shell,
assuming $\Omega_{\nu} =0.005$ and the other cosmological parameters as in Table \ref{table:fiducial}. The curves are normalized to agree 
at $l=1$. The envelope of error bars is derived from Eq. \ref{eq:sigmacl}.
}.  
\label{fig:errclnl}
\end{figure}

\begin{figure}
\includegraphics[width=8cm,height=5cm]{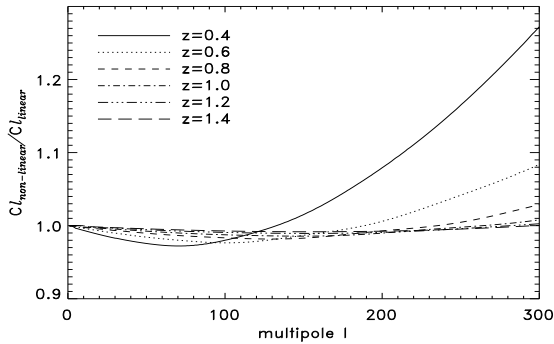}
\caption{ The ratio of $C_l$'s with non-linear to linear power spectra according to Eq. \ref{eq:pknl} including epoch dependence (solid line) for 4 shells.
assuming $\Omega_{\nu} =0.005$ and the other cosmological parameters as in Table \ref{table:fiducial}. The curves are normalized to agree 
at $l=1$. We see that up to $l=100$ the effect of non-linear power spectrum is less than 5 \%.}. 
\label{fig:clnl}
\end{figure}

For comparison we plot in Figure \ref{fig:errclnl}  the errors on the $C_{\ell}$'s \citep[e.g.][]{dod03,blk07}:
\begin{equation}
\sigma(C_{\ell}) = \sqrt{(2/f_{\rm sky}/(2 l+1)} (C_{\ell} + \frac{1}{N/\Delta\Omega}),  
\label{eq:sigmacl}
\end{equation}
which are far larger than the non-linearity effect for $\ell < 100$.
As Figure \ref{fig:clnl} shows the that in this regime the effect of non-linearity is  less than 5 \%. 

\subsection{Epoch dependent biasing}

Another source of uncertainty in constraining neutrino mass from
galaxy surveys is galaxy biasing \citep[e.g.][]{elg05}.
The bias systematically increases with redshift for two reasons:
\begin{enumerate}
\item In standard models of the evolution of galaxy clustering, the
  bias factor of a class of galaxies increases with redshift in
  opposition to the decreasing linear growth factor, in order to
  reproduce the observed approximate constancy of the small-scale
  clustering length \citep[e.g.][]{mag00,lah02}.
\item In a flux limited survey galaxies in more distant redshift slices are preferentially more
  luminous (owing to the fixed apparent magnitude threshold) and hence
  more strongly clustered \citep{norb02}.  
\end{enumerate}

Here we test what happens if the biasing varies with epoch.
One simple model is the `galaxy evolving model' \citep{fry96},
where the bias evolves as:
\begin{equation}
b(z) = 1 + (b_0 -1)/D(z)
\label{eq:bz}
\end{equation}
where $b_0$ is the present galaxy bias for a particular galaxy type,
and $D(z)$ is the linear theory growth rate.

 The sensitivity to the assumed galaxy biasing scheme is illustrated in Figure \ref{fig:contsig8} (bottom).   
 The point  $O$ corresponds to $b=1.00$ in the fiducial model for all shells,  $X$ corresponds to $b=1.02$, and $Y$ corresponds to epoch-dependent $b(z)$ from Eq. \ref{eq:bz} with $b_0=1.02$. 
 According to that plot, if $b$ deviates from unity by more than $2\%$ then $\sigma_8$ changes by $2\%$ and $\Omega_{\nu}$ changes by $7\%$. Therefore biasing is the most sensitive quantity in our analysis.
In other words, we need to have the bias known to 2 per cent accuracy as a larger bias would introduce a best fit value outside the one sigma error bar in figure 4.

We see that our results are sensitive to biasing.
Fortunately there are several independent ways of controlling
this systematic effect : (i) modelling the biasing via halo model or
semi-analytic simulations, and marginalising over the biasing
parameters; (ii) estimating the neutrino mass from galaxy
power-spectra derived for different galaxy types and checking for
consistency; (iii) estimating the biasing empirically from
weak-lensing map to be produced e.g. from DES itself;
(iv) estimating biasing from high order statistic such as the bi-spectrum.

\section{Conclusions}
\label{sec:conclu}

We study the prospects for detecting neutrino masses from the galaxy
angular power spectrum in photo-z shells in the Dark  Energy Survey,  combined with CMB fluctuations as will be measured by Planck.
Although the core science case for DES is Dark Energy, we see that DES can provide us with other important extra science,
such as neutrino mass. Our main conclusions are:

\begin{itemize}

\item We forecast for DES\&Planck a 2-sigma error of total neutrino mass $\Delta M_{\nu}
\approx 0.12$ eV.  If the true neutrino mass is very
close to zero, then we can obtain an upper limit of 0.11 eV (95\% CL).             

\item This upper limit  from DES+Planck is over  3 times tighter than 
using Planck alone, as DES breaks the parameter degeneracies in a CMB-only analysis.

\item The results are sensitive to the assumed galaxy biasing, and stand if the galaxy bias in known to within 2 per cent. This is feasible given other analyses of the galaxy bias such as the three point correlation function \citep{ross07}.

\item The results are robust  to uncertainties in non-linear fluctuations and redshift distortion. 

\item The results  are similar  if we  supplement DES bands (grizY) with the
VISTA Hemisphere Survey (VHS) near infrared band (JHK).

\end{itemize}

DES can also be used to extract information on neutrino mass via other techniques, e.g. 
weak gravitational lensing, as considered recently for other imaging surveys  \citep{kit08, ichi09}.
We note that the  level of sensitivity for neutrino mass from DES\& Planck is of much
relevance for comparison with the direct measurement of the neutrino mass from
laboratory experiments. E.g. the KATRIN tritium beta decay
experiment. Furthermore, the DES \& Planck measurements can be combined with
laboratory experiments to derive more accurate neutrino masses
\citep{host07}.

\section*{Acknowledgments}

We thank our DES collaborators  for helpful  discussions,  S. Saito, M. Takada \&  A. Taruya  for 
providing us with a code for a non-linear power spectrum, and S. Bridle, J. Thaler  and S. Thomas for comments on the manuscript.
OL acknowledges a Royal Society Wolfson Research Merit Award and an
Erna \& Jakob Michael Visiting Professorship at  the Weizmann Institute of Science, 
AK acknowledges a Perren Studentship
and  FBA acknowledges a Leverhulme Early Career Fellowship.


\begin{thebibliography}{}
\bibitem[Abdalla \& Rawlings(2007)]{abd07} Abdalla F.~B., \& Rawlings S.,2007, MNRAS, 381, 1313 
\bibitem[Abdalla et al.(2009)]{abd09} Abdalla F.~B., Blake C., \& Rawlings S.,2009, arXiv: 0905.4311
\bibitem[Banerji et al.(2008)]{manda08} Banerji M., Abdalla F.~B., Lahav O., \& Lin H.,2008, MNRAS, 386, 1219 
\bibitem[Blake \& Bridle(2005)]{blk05} Blake C., \& Bridle S.,2005, MNRAS, 363, 1329 
\bibitem[Blake et al.(2007)]{blk07} Blake C., Collister A., Bridle S., \& Lahav O.,2007, MNRAS, 374, 1527
\bibitem[Bucher et al.(2002)]{buch02} Bucher M., Moodley K., \& Turok N.,2002, PRD, 66, 023528
\bibitem[Collister \& Lahav(2004)]{coll04} Collister A., \& Lahav O.,2004,PASP, 116, 345
\bibitem[Dodelson(2003)]{dod03} Dodelson S.,2003, Modern cosmology,Academic Press.
\bibitem[Dolney et al.(2004)]{dol04} Dolney D., Jain B., \& Takada M.,2004, MNRAS, 352, 1019
\bibitem[Elgar{\o}y et al.(2002)]{elg02} Elgar{\o}y {\O}. et al.,2002, Physical Review Letters, 89, 061301
\bibitem[Elgar{\o}y \& Lahav(2005)]{elg05} Elgar{\o}y {\O}. \& Lahav O.,2005, New Journal of Physics, 7, 61 
\bibitem[Fisher et al.(1994)]{fish94} Fisher K.~B., Scharf C.~A. \& Lahav O.,1994, MNRAS, 266, 219
\bibitem[Fry(1996)]{fry96} Fry J.~N., 1996, APJL, 461, L65
\bibitem[Hannestad et al.(2006)]{hann06} Hannestad S., Tu H., \& Wong Y.~Y.,2006, Journal of Cosmology and Astro-Particle Physics, 6, 25
\bibitem[Host et al.(2007)]{host07} Host O., Lahav O., Abdalla F.~B., \& Eitel K.,2007, PRD, 76, 113005
\bibitem[Hu et al.(1998)]{hu98} Hu W., Eisenstein D.~J., \& Tegmark M.,1998, Physical Review Letters, 80, 5255
\bibitem[Ichiki et al.(2009)]{ichi09} Ichiki K., Takada M., \& Takahashi T.,2009, PRD, 79, 023520
\bibitem[Kiakotou et al.(2008)]{kia08} Kiakotou A., Elgar{\o}y {\O}., \& Lahav O.,2008,PRD, 77, 063005
\bibitem[Kitching et al.(2008)]{kit08} Kitching T.~D., Heavens A.~F., Verde L., Serra P., \& Melchiorri A.,2008,PRD, 77, 103008 
\bibitem[Komatsu et al.(2009)]{koma09} Komatsu E., et al.,2009, APJS, 180, 330
\bibitem[Lahav et al.(2002)]{lah02} Lahav O., et al.,2002, MNRAS, 333, 961
\bibitem[Lesgourgues \& Pastor(2006)]{lesg06} Lesgourgues J., \& Pastor S.,2006, PhysRep, 429, 307 
\bibitem[Lewis et al.(2000)]{lew00} Lewis A., Challinor A., \& Lasenby A.,2000, APJ, 538, 473
\bibitem[Magliocchetti et al.(2000)]{mag00} Magliocchetti M., Bagla J.~S., Maddox S.~J., \& Lahav O.,2000,MNRAS, 314, 546
\bibitem[Norberg et al.(2002)]{norb02} Norberg P., et al.,2002, MNRAS, 336, 907
\bibitem[Padmanabhan et al.(2007)]{pad07} Padmanabhan N., et al.,2007, MNRAS, 378, 852
\bibitem[Peebles(1973)]{peeb73} Peebles P.~J.~E.,1973, APJ, 180, 1
\bibitem[Ross et al. (2007)]{ross07} Ross, A.J., Brunner,  R.J., Myers, A.D., 2007, APJ, 665, 67
\bibitem[Saito et al. (2008)]{sait08} Saito S., Takada M., \& Taruya A.,2008, Physical Review Letters, 100, 191301
\bibitem[Saito et al. (2009)]{sait09} Saito S., Takada M., \& Taruya A.,2009, arXiv:0907.2922
\bibitem[Scharf et al.(1992)]{sch92} Scharf C., Hoffman Y., Lahav O., \& Lynden-Bell D.,1992, MNRAS, 256, 22
\bibitem[Seo \& Eisenstein(2003)]{seo03} Seo H.-J., \& Eisenstein D.~J.,2003, APJ, 598, 720
\bibitem[Smith et al.(2003)]{smit03} Smith R.~E., et al.,2003, MNRAS, 341, 1311
\bibitem[Wandelt et al.(2001)]{wand01} Wandelt B.~D., Hivon E., \& G{\'o}rski K.~M.,2001, PRD, 64, 08300
\bibitem[Wright et al.(1994)]{wri94} Wright E.~L., Smoot G.~F., Bennett C.~L., \& Lubin P.~M.,1994, APJ, 436, 443 
\bibitem[Zhan et al.(2006)]{zhan06} Zhan H., Knox L., Tyson J.~A., \& Margoniner V.,2006, APJ, 640, 8

\end{thebibliography}
\end{document}